\documentclass[aps,showpacs,prb,superscriptaddress,floatfix,twocolumn,citeautoscript]{revtex4-1}
\usepackage{amstext}
\usepackage{graphicx}
\usepackage{amsmath}
\usepackage{amssymb}
\usepackage{bm}
\usepackage{dcolumn}
\usepackage{subfigure}
\usepackage{units}
\usepackage{hyperref}
\usepackage[usenames]{color}
\usepackage{booktabs}
\usepackage{multirow}
\usepackage{siunitx}
\usepackage{wrapfig}
\usepackage{lipsum}
\usepackage{setspace}
\hypersetup{pdfborder=0 0  0,colorlinks=true,citecolor=blue,linkcolor=blue}
\input epsf
\newcommand{\BN}{\textit{h}-BN}

\newcommand{\MS}{MoS$_2$}

\begin{document}
	
	\title{Plasmon modes in monolayer and double-layer black phosphorus under applied uniaxial strain} 
	
	\author{S. Saberi-Pouya}
	\affiliation{Department of Physics, Shahid Beheshti University, G.C., Evin, Tehran 1983969411, Iran}
	\affiliation{Department of Physics, University of Antwerp, Groenenborgerlaan 171, B-2020 Antwerpen, Belgium}
	\author{T. Vazifehshenas}
	\email{t-vazifeh@sbu.ac.ir}
	\affiliation{Department of Physics, Shahid Beheshti University, G.C., Evin, Tehran 1983969411, Iran}
	\author{M. Saleh}
    \affiliation{Department of Physics, Shahid Beheshti University, G.C., Evin, Tehran 1983969411, Iran}
	\author{M. Farmanbar}
	\affiliation{Faculty of Science and Technology and MESA+ Institute for Nanotechnology, University of Twente, P.O. Box 217, 7500 AE Enschede, The Netherlands}
	\author{T. Salavati-fard}
	\affiliation{Department of Physics and Astronomy, University of Delaware, Newark, DE 19716, USA}
	
	\begin{abstract}
We study the effects of an applied in-plane uniaxial strain on the plasmon dispersions of monolayer, bilayer and double-layer phosphorene structures in the long-wavelength limit within the linear elasticity theory. In the low energy limit, these effects can be modeled through the change in the curvature of the anisotropic energy band along the armchair and zigzag directions. We derive analytical relations for the plasmon modes under uniaxial strain and show that the direction of the applied strain is important. Moreover, we observe that along the armchair direction, the changes of the plasmon dispersion with strain  are different and larger than those along the zigzag direction. Using the analytical relations for two-layer phosphorene systems, we find that the strain-dependent orientation factor of layers could be considered as a means to control the variations of the plasmon energy. Furthermore, our study shows that the plasmonic collective modes are more affected when the strain is applied equally to the layers compared to the case in which the strain is applied asymmetrically to the layers. We also calculate the effect of strain on the drag resistivity in a double-layer phosphorene structure and obtain that the changes in the plasmonic excitations, due to an applied strain, are mainly responsible for the predicted results. This study can be easily extended to other anisotropic two-dimensional materials.
	\end{abstract}
	\date{\today}
	\maketitle

\section{Introduction}
	
Black phosphorus (BP) is the most stable allotrope form of crystalline phosphorus and is a layered van der Waals (vdW) material like graphite. BP is a semiconductor with puckered orthorhombic structure offering highly anisotropic optical and electronic properties\cite{Samira:Phonon2017,Xia:nat13,0957-4484-27-5-055701,SOC:Samira,liu:nano15}. Few-layer BP has a unique feature that its direct bandgap exponentially deceases from ~1.5 to 0.35 eV as the number of layers increases from monolayer to its bulk form. This property allows layer-engineering to tune the electronic bandgaps and light absorption spectra of BP. 

This structure, with no surface dangling bonds withstands high deformation without breaking. Strain engineering, a recently developed and widely adopted technique, is capable of changing the electronic and optical properties of BP. The band structure of this novel material is highly sensitive to an applied strain and also deformation\cite{0957-4484-26-7-075701}, making it a good potential candidate for electromechanical applications\cite{PhysRevB.90.205421,mohammadi2016strain}. While silicon typically breaks at strain level of 1.5 \% \cite{Mohan201440} and \MS \ gets folded and wrinkled by a tensile strain of about 10 \%  \cite{Shi:prb13,WonSeok:prb85}, monolayer BP (phosphorene) is able to withstand a surface tension and tensile strain up to 10 N/m and 30 \%, respectively\cite{PhysRevB.90.205421,1367-2630-17-3-035008,0957-4484-26-7-075701,PhysRevB.94.085417,PhysRevB.90.085402,0953-8984-27-17-175006}.

Plasmons, the collective excitations of the oscillating charges have been studied in graphene and other two-dimensional (2D) materials, extensively \cite{RevModPhys.54.437, 0953-8984-21-2-025506, PhysRevB.87.235418, fb4ef8900275464a9b3996b3aa006d78}. 
	It was shown that plasmonic collective modes in graphene have relatively long propagation length \cite{Constant:nat16,GraphenePlasmonics}. In a doped \MS\ monolayer, as a result of a strong spin-orbit coupling, the plasmon modes are different from those in graphene-like materials and enter the electron-hole continuum region just similar to the case of 2D electron or hole gas with spin-orbit coupling\cite{PhysRevB.83.115135,PhysRevB.87.085321}. The effect of uniaxial strain on the dispersion relation of plasmons in graphene was investigated and a strain-induced anisotropic enhancement of the deviations from linearity of the transverse modes was obtained \cite {PhysRevB.84.195407}. Also, the anisotropic collective excitations in pristine single-layer and multi-layer BP have been studied in the absence and presence of a magnetic or electric field \cite{PhysRevB.92.115440,PhysRevB.92.085408}. In the context of plasmons in layered BP, the effect of strain is one of the main issues that should be addressed. Particularly, it is interesting to explore the strain influence on the spatially separated double-layer phosphorene system in which the layers are coupled only via the long-range Coulomb interaction and the system exhibits a rich variety of phenomena like Coulomb drag \cite{RevModPhys.88.025003}. The interaction-induced Coulomb drag transresistivity in phosphorene double-layer, that was recently investigated \cite{Samira:drag2016}, is attractive because of its dependence on the plasma oscillations and sensitivity to the structural anisotropy of the system.	

In this paper, we theoretically investigate the effects of an applied uniaxial strain on the plasmon dispersion of monolayer, double-layer and bilayer phosphorene structures within the linear elasticity theory. We calculate the strain-dependent dynamic dielectric function within the random phase approximation (RPA) and obtain the plasmon dispersion relation in the long-wavelength limit.
 We show that the energy of plasmon modes depends upon the applied strain through the resulting effective masses along the armchair ($x$-axis) and zigzag ($y$-axis) directions. In two-layer phosphorene systems, the strain-dependent orientation factor of layers control the variations of plasmon frequencies. We also present our numerical results for the strain effects on the drag resistivity in a double-layer phosphorene structure and conclude that the changes in plasmonic excitations, due to applied strain, are mainly responsible for the predicted behaviors. 

This work is organized as follows: In Sec.\ref{theory1}, we introduce
the low-energy Hamiltonian and effective masses for the monolayer phosphorene under strain. In Sec. \ref{theory2}, the temperature-dependent polarization function, the RPA formalism for the dielectric function and the long-wavelength plasmon spectra for the monolayer and double-layer cases are presented and discussed. The effect of applied uniaxial strain on the Coulomb drag as an example of transport properties of double-layer phosphorene is studied in Sec. \ref{theory3}. Finally, in Sec.\ref{conclusion} we summarize the main results of this paper.

\section{ Strain-dependent electronic structure and effective mass of phosphorene in the linear deformation regime}
\label{theory1}
 We consider a system composed of the phosphorene layers under externally applied strain.

  \begin{table}[hb]
  	\caption{Physical parameters of phosphorene under an uniaxial strain along the armchair ($x$-axis) and zigzag ($y$-axis) edges taken from Ref$\cite{mohammadi2016strain}$.}

  	\begin{ruledtabular}
  		\begin{tabular}{c c c  } 
  			
  			& \qquad\quad $x$-axis  &  $y$-axis \\ [0.5ex]  
  			
  			\hline 
  			
  			\\
  			$u^{\prime}$ (eV) &\quad\quad 0.33 & 0.19 \\
  			$\delta^{\prime}$ (eV) &\quad\quad 2.15 & 2.91 \\
  			\\
  			$\chi^{\prime}$ (eV$\mathring{A}$) &\quad\quad -2.96 & 3.43  \\
  			\\
  			$\eta_x^{\prime}$ (eV$\mathring{A}^2$)   &\quad\quad 1.21 &-0.45 \\
  			$\eta_y^{\prime}$ (eV$\mathring{A}^2$)   &\quad\quad -0.45  & 0.89  \\
  			$\gamma_x^{\prime}$ (eV$\mathring{A}^2$) &\quad\quad 2.37  & -3.81  \\
  			$\gamma_y^{\prime}$ (eV$\mathring{A}^2$) &\quad\quad -3.81  & 3.81  \\
  		  		\end{tabular}
  		\label{table:nonlin} 
  	\end{ruledtabular}
  \end{table}
  
    The unit cell of monolayer phosphorene contains four phosphorus atoms which are stacked in puckered subplanes. In the tight-binding (TB) model, the Hamiltonian of the system is given as \cite{PhysRevB.89.201408}
  $ H_t=\sum_{i,j} t_{ij} c^{\dagger}_i(c_j) $,
  \noindent where $ t_{ij} $ represents hoping between $i$th and $j$th sites and $c^{\dagger}_i (c_j)$ is the creation (annihilation) operator of electrons at site $i(j)$. A strain-dependent two-band TB Hamiltonian in the continuum approximation (retaining the terms up to second order in $k$) and linear deformation regime has been obtained around the $\Gamma$ point as \cite{PhysRevB.92.075437,mohammadi2016strain}

\begin{widetext}
\begin{equation}
\hat{H}^s_{\textbf{k}}=\left(\
\begin{array}{cc}
u^s+\eta^s_xk^2_x+\eta^s_yk^2_y&\delta^s+\gamma^s_xk^2_x+\gamma^s_yk^2_y+i
\chi^s k_x\\
\delta^s+\gamma^s_xk^2_x+\gamma^s_yk^2_y-i\chi^s k_x&u^s+\eta^s_xk^2_x+\eta^s_yk^2_y
\end{array}\right).
\label{1}
\end{equation}
\end{widetext}

 Here, only the uniaxial strains $s_x$, $s_y$ and $s_z$ along the three principle directions, namely, the $x$-axis, $y$-axis and also the direction normal to the phosphorene plane ($z$-axis) have been considered. Therefore, the changes of position vector components are determined by $\mathbb{X}^{s}=(1+s_{\mathbb{X}})\mathbb{X}$ where $\mathbb{X}=x,y$ and $z$. The strain-dependent parameters in the above TB Hamiltonian are related to the unstrained ones as $ u^{s}=u+s u'$,  $\eta^{s}_x=\eta_x+s {\eta}'_x$, $\eta^{s}_y=\eta_y+s {\eta}'_y$, $\gamma^{s}_x=\gamma_x+s {\gamma}'_x$, $\gamma^{s}_y=\gamma_y+s {\gamma}'_y$ and $\chi^{s}=\chi+s {\chi}'$ where $u=0.42$ eV, $\eta_x=1.03$ eV$\mathring{A}^2$, $\eta_y=0.56$ eV$\mathring{A}^2$, $\delta=0.76$ eV, $\gamma_x=3.51$ eV$\mathring{A}$, $\gamma_y=3.81$ eV$\mathring{A}^2$ and $\chi=-5.34$ eV$\mathring{A}$. In Table \ref{table:nonlin}, we list the calculated parameters of the TB Hamiltonian matrix elements for a phosphorene monolayer under the uniaxial strain\cite{mohammadi2016strain}.  

 \begin{figure*}[ht]
 	\includegraphics[width=18.0cm]{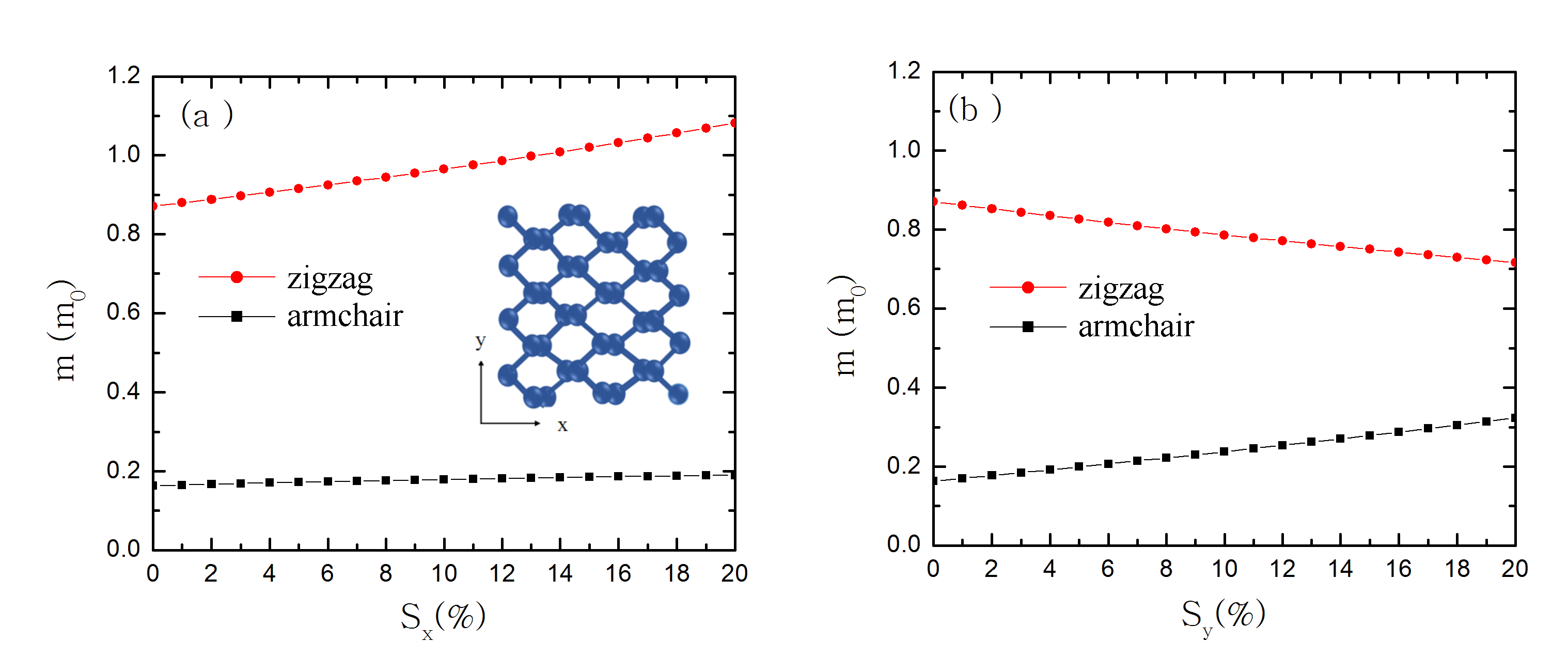}
 	\caption{Effective masses along the armchair and zigzag directions of phosphorene as functions of a tensile strain applied along (a)$x$-axis (s$_x$) and (b) $y$-axis (s$_y$).}
 	\label{fig:1}
 \end{figure*}
Applied strain can, in fact, affect the inter-band coupling by changing the energy gap and $\chi^{s}$. However, the relatively large bandgap of the phosphorene monolayer which results in a weak inter-band interaction, allows one to decouple the conduction and valence bands at small wave vectors. In this approximation, the phosphorene band structure is obtained as:
\begin{equation}
E^s_{\pm}({\textbf{k}})\approx u^s+\eta^s_x k^2_x +\eta^s_y k^2_y \pm \left(\delta^s+\left[{\gamma^s_x}k^2_x +{\gamma^s_y}k^2_y +\dfrac{(\chi^s)^2}{2\delta^s} k^2_x\right]\right),
\label{6}
\end{equation}
where +(-) stands for the conduction (valence) band. Thus, the strain-dependent electron and hole effective masses are given by
 \begin{equation}
 	\begin{aligned}
 		&m^{s}_{ex}=\dfrac{{\hbar^2}}{2(\eta^{s}_x+\gamma^{s}_x+(\chi^{s})^2/2\delta^{s})}\\
 		&m^{s}_{ey}=\dfrac{{\hbar^2}}{2(\eta^{s}_y+\gamma^{s}_y)}\\
 		&m^{s}_{hx}=\dfrac{{\hbar^2}}{2(\gamma^{s}_x-\eta^{s}_x+(\chi^{s})^2/2\delta^{s})}\\
 		&m^{s}_{hy}=\dfrac{{\hbar^2}}{2(\gamma^{s}_y-\eta^{s}_y)}. 
 	\end{aligned}\label{16}
 \end{equation} 
 Considering only the conduction band and using the above strain-dependent electron effective masses along the armchair ($m^{s}_{x}=m^{s}_{ex}$) and zigzag directions ($ m^{s}_{y}=m^{s}_{ey}$), we can obtain the following expression for the energy of electrons\cite{Low:prl14,PhysRevB.92.075437}
 \begin{equation}
 	E^s(\mathbf{k})=\frac{\hbar^2}{2}(\frac{k_{x}^{2}}{m^s_{x}}+\frac{k_{y}^{2}}{m^s_{y}}).
 	\label{eq:energy}
 \end{equation}

The calculated strain-dependent electron effective masses of phosphorene as functions of $s_x$ and $s_y$ are depicted in Figs. \ref{fig:1}(a) and (b). The inset figure is the top view of monolayer phosphorene. In a relaxed (unstrained) phosphorene monolayer system, the electron effective mass is predicted to be $0.168$m$_0$ in the armchair and $0.852$m$_0$ in the zigzag directions \cite{PhysRevB.89.201408,mohammadi2016strain}. The smaller electron effective mass, about an order of magnitude, along the armchair direction results in a favorable transport direction in phosphorene.
We show that with applying the uniaxial strain along $x$-axis, the effective mass in the armchair direction is almost insensitive to the strength of applied strain while along the zigzag direction, it increases notably with strain. These behaviors are direct consequences of the strain effect on the band structure and atomic orbitals of phosphorene \cite{PhysRevB.90.085402}. In the case of strain applied parallel to the $y$-axis, the effective mass along the armchair (zigzag) direction increases (decreases). Since we just consider the one-band model, with no surprise, there is not any sharp changes in the effective masses due to the direct-indirect bandgap transition (see Fig. \ref{fig:1}) \cite{PhysRevB.90.085402}.

\section{Plasmons in monolayer and double-layer phosphorene systems under uniaxial strains}
\label{theory2}                           
In order to study the plasmonic collective modes, we need to calculate the dielectric function $ \epsilon(q, \omega)$ of the system. Within the linear-response theory, plasmon modes are defined as the zeros of the dielectric function. The RPA is the simplest diagrammatic procedure to include the electron correlations in the dielectric function. It is well-known that RPA gives the exact dielectric function at the limit of infinite electron density. As a result, it makes sense to employ the RPA to calculate the dielectric function at high electron density. The strain-dependent RPA dielectric function matrix for a few-layer system in the absence of inter-layer tunneling is given by \cite{Hwang:prb09,Wen:nanotech12}

\begin{equation}
\epsilon^s_{ij}(\mathbf{q},\omega)=\delta_{ij}+V_{ij}(q)\Pi^s_{i}(\mathbf{q},\omega),
\label{eq:epsil}
\end{equation}

\noindent where $V_{ij}(q)=\nu(q)\exp(-qd(1-\delta_{ij}))$ is the unscreened 2D Coulomb interaction with $d$ being the separation between layers and $\nu(q)=2\pi e^{2}/q\kappa$, where $\kappa$ being the average dielectric constant. The non-interacting polarization function of $i$th layer can be obtained through the following equation\cite{Bohm:pr53}:
\begin{equation}
\Pi^s_{i}(\mathbf{q},\omega)=-\frac{2}{A} \sum \limits_{\mathbf{k}} \frac{f^{0}(E^s_{i}(\mathbf{q}))-f^{0}(E^s_{i}(\mathbf{k+q}))}{E^s_{i}(\mathbf{q})-E^s_{i}(\mathbf{k+q})+\hbar\omega+i\xi}. 
\label{eq6}
 \end{equation}

\begin{figure}[ht]
	\includegraphics[width=8.3cm]{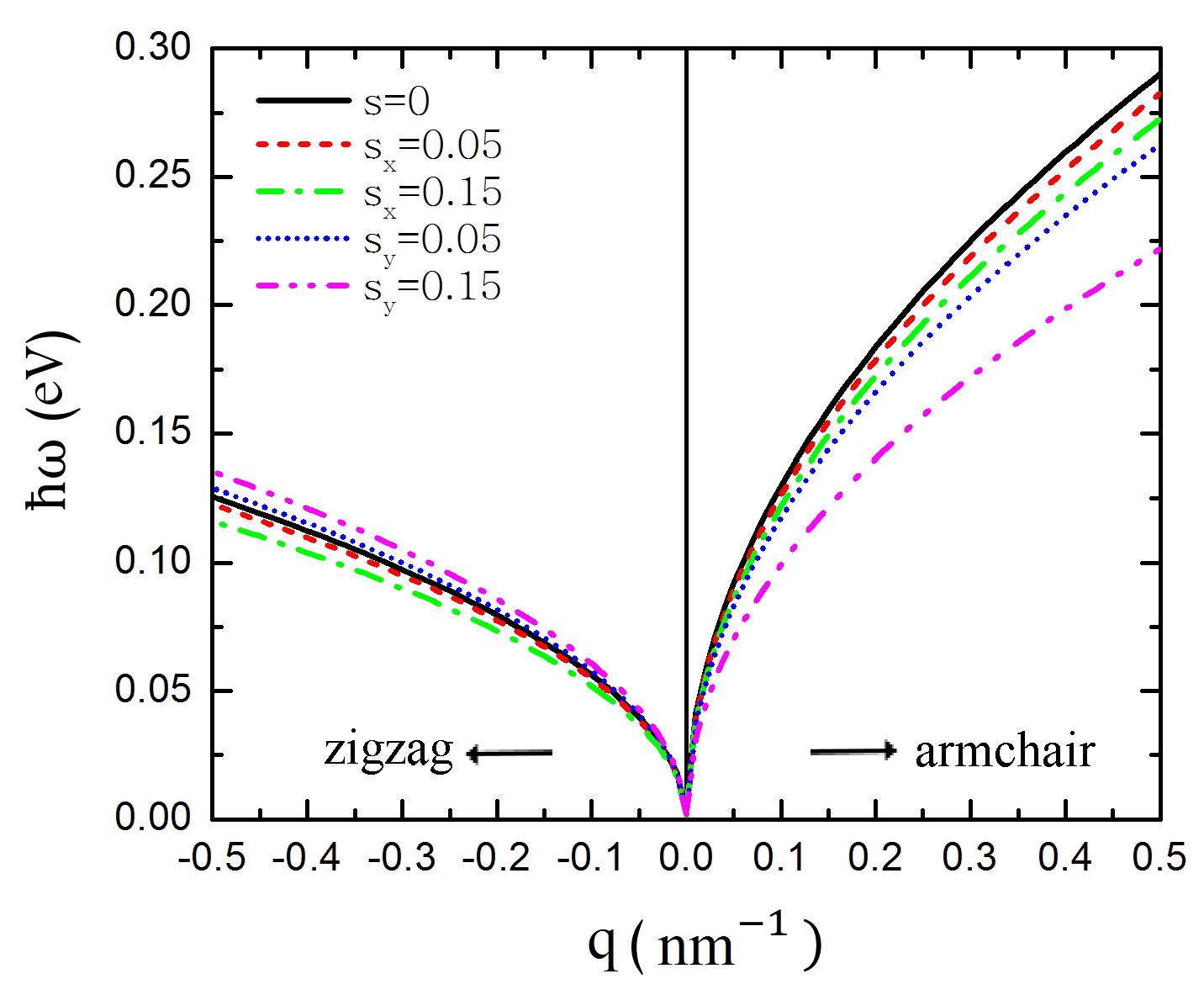}
	\caption{The plasmon dispersions in the unstrained and uniaxial tensile strained phosphorene monolayers along $x-$ and $y-$axis with $n= 1\times10^{13}$cm$^{-2}$. Here, {\BN} is considered as the substrate.}
	\label{fig:2}
\end{figure}

\noindent Here, $A$ denotes the area of the unit cell, $f^{0}(E^s_{i}(\mathbf{q}))$ is the Fermi distribution function of layer $i$ at strain-modified energy $E^s_{i}$ corresponding to a 2D wave vector $\mathbf{q}$ and $\xi$ being the broadening parameter which accounts for disorder in the system. In a phosphorene monolayer, the temperature-dependent dynamic polarization function for the intra-band transition has been calculated by making use of the anisotropic parabolic energy dispersion relation (Eq. (\ref{eq:energy})) as
\cite{Low:prl14,Samira:drag2016}:

\begin{equation}
\begin{aligned}
\frac{\Pi^s_{i}(\mathbf{q},\omega)}{g^s_{2d}}={}&-\int dK \frac{\Phi^s_{i}(K,T)}{Q}\Bigg[sgn(\Re(Z_{-}))\frac{1}{\sqrt{Z_{-}^{2}-K^{2}}}\\
&-sgn(\Re(Z_{+}))\frac{1}{\sqrt{Z_{+}^{2}-K^{2}}}\Bigg].
\label{eq:10}
\end{aligned}
\end{equation}
In the above symmetric form of finite temperature anisotropic polarization function, we have defined $\mathbf{K(Q)}=\sqrt{m^{s}_{d}/\hat{M^s}}(\mathbf{k(q)}/k_{F})$, where $\hat{M^s}$ is the strain-dependent mass tensor with diagonal elements $m^s_{x}$ and $m^s_{y}$ and $m^{s}_{d}=\sqrt{m^{s}_{x}m^{s}_{y}}$ is the 2D density of states mass. Here, $k_{F}=\sqrt{2\pi n}$ is the Fermi wave vector, $n$ being the electron density, $\mathbf{q}=q(\cos\theta,\sin\theta)$, $g^s_{2d}=m^{s}_{d}/\pi \hbar^{2}$ and the rotational angle, $\tau_{i}$, is defined as the angle between $x$-axis in the laboratory frame and $x$ direction of the $i$th layer. So, we can write $Q=q\sqrt{m^{s}_{d}R^{s}_{i}(\theta)}/k_{F}$ in which the strain-dependent orientation factor, $R^{s}_{i}(\theta)$, is expressed as:

\begin{gather}
R^{s}_{i}(\theta)=\bigg(\frac{\cos^{2}(\theta-\tau_{i})}{m^{s}_{x}}+\frac{\sin^{2}(\theta-\tau_{i})}{m^{s}_{y}}\bigg).
\label{eq:R}
\end{gather} 

Moreover, in Eq. (\ref{eq:10}) we define $Z_{\pm}=((\hbar\omega+i\xi)/\hbar Q \nu^s_{F})\pm(Q/2 k_F)$ with $\nu^s_{F}=\hbar k_{F}/m^{s}_{d}$ and $\Phi^s_{i}$ is given by:
\begin{equation}
\Phi^s_{i}(K,T)=\frac{K}{1+\exp[(K^{2}E^s_{F,i}-\mu^s_{i})/k_{B}T]},
\end{equation}
where  $E^s_{F,i}$ and $\mu^s_{i}$ are the strain-dependent Fermi energy and chemical potential of layer $i$, respectively and satisfy the following particle number conservation condition \cite{Ashcroft}:
\begin{equation}
E^s_{F,i}=\mu^s_{i}+k_{B}T\ln[1+\exp(-\mu^s_{i}/k_{B}T)].
\label{eq:chem}
\end{equation}

In the limit of long-wavelength where the plasmon excitations have relatively long lifetimes, the zero-temperature polarization function can be approximated as \cite{Samira:Phonon2017}:   
\begin{equation}
\frac{\Pi^s_{i}(\omega,q,\theta)}{g^s_{2d}}\approx \frac{q^{2}}{\omega^{2}}R^s_{i}(\theta)E^s_{F,i} .
\label{eq:polq}
\end{equation}
  
\noindent Thus, the long-wavelength plasmon dispersion relation of a phosphorene monolayer under uniaxial strain is obtained as
  
\begin{equation}
\omega^s_{pl}(q,\theta)=\sqrt{\frac{2n\pi e^{2}R^s_{i}(\theta)q}{\kappa}},
\label{Wm}
\end{equation}

\begin{figure*}[ht]
	\includegraphics[width=18.0cm]{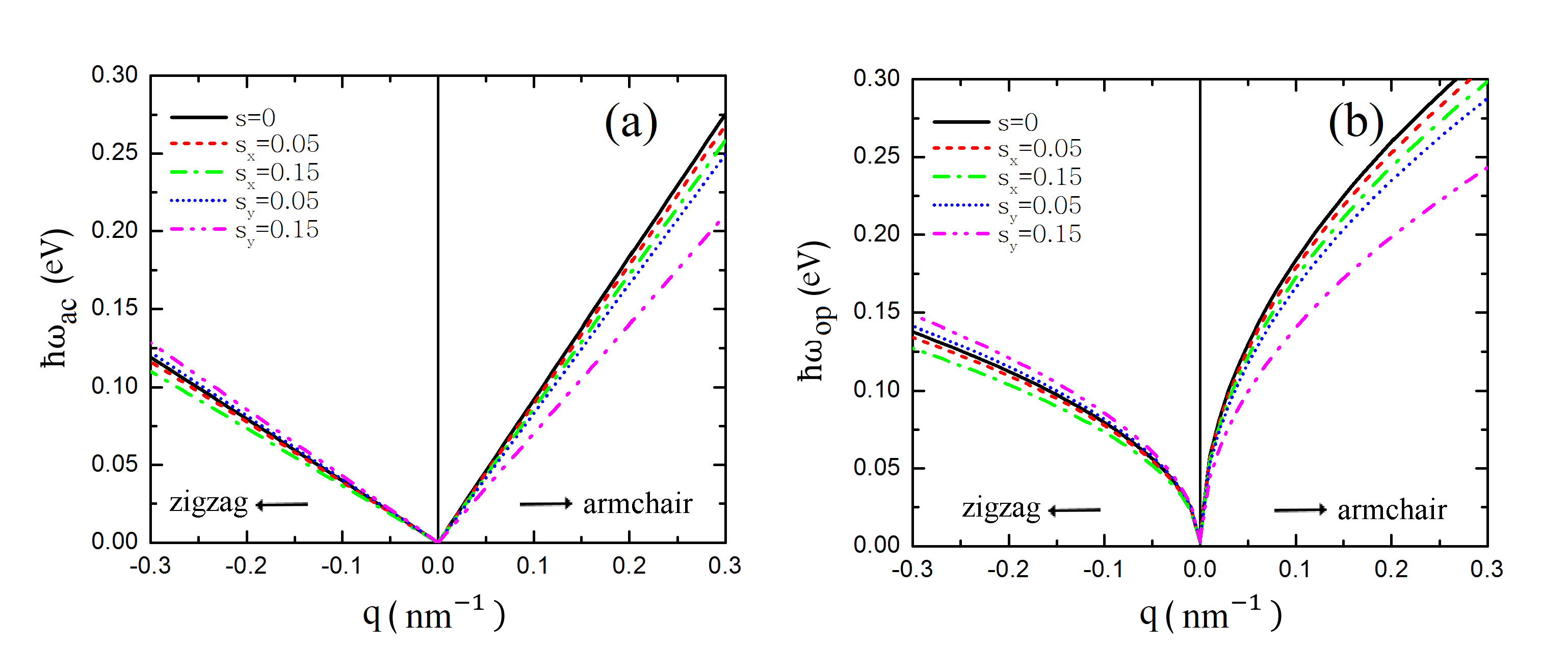}
	\caption{(a) Acoustic and (b) optical plasmon dispersions along armchair and zigzag directions of a double-layer phosphorene structure. Uniaxial tensile strain is applied along the $x-$ and $y$-axis with $n= 1\times10^{13}$cm$^{-2}$ and $d$ =5 nm. The structure is sandwiched by \BN.}
	\label{fig:3}
\end{figure*} 
  
\noindent which depends on strain through the orientation factor. An important and useful feature of this formulation is that all the effects due to strain appear in the strain-dependent orientation factor $R^s_i(\theta)$ as a multiplicative modifier. Therefore, we recover the typical $\sqrt{q}$ dispersion of anisotropic 2D plasmons and the strain effects are included in the effective masses. In Fig. \ref{fig:2}, we show our numerical results for the strain-dependence plasmon modes in monolayer phosphorene along (a) armchair and (b) zigzag directions for a few applied uniaxial strains. For comparison, we also present numerical results for the case of relaxed system. Due to the peculiar behaviors of the effective masses under uniaxial strains along zigzag and armchair directions, plasmon modes behave differently along these directions. The energy of plasmon modes, $\hbar\omega^s_{pl}(q,\theta)$, along the armchair direction decreases with increasing the strength of applied strain due to effective mass enhancement along this direction. However, as it is shown in Fig. \ref{fig:2}, the effect of applied strain on the plasmon dispersion along the zigzag direction is different. It is important to notice that the energy of plasmon modes in the zigzag direction decreases (increases) corresponding to the applied $s_x$ ($s_y$) strain. This observation could be understood through the fact that the electron effective mass increases (decreases) with increasing $s_x$ ($s_y$) along the zigzag direction (see Fig. \ref{fig:1}). However, the effective mass along the armchair direction increases with increasing $s_x$ and also $s_y$. It is worth pointing out that while the applied strain changes the energy of plasmon modes, it does not alter the standard $\sqrt{q}$ dispersion behavior. 
    
To show how unaxial strain affects the plasmonic excitations in double-layer phosphorene, we solve Eq. (\ref{eq:epsil}) for a 2$\times$2 matrix. Considering the long-wavelength limit and only the intra-band transition, there are two plasmonic branches, the so-called acoustic plasmons with $q$-linear behavior 
     
  \begin{equation}
  \omega^s_{ac}(q,\theta)=2q\sqrt{\frac{n\pi e^2d}{\epsilon_{\infty}}\bigg(\frac{R^s_{1}(\theta)R^s_{2}(\theta)}{R^s_{1}(\theta)+R^s_{2}(\theta)}\bigg)},
  \label{eq:Wc}
  \end{equation}  
 and optical plasmons with square-root dependence on $q$  
   
 \begin{equation}
 \omega^s_{op}(q,\theta)=\sqrt{\frac{2n\pi e^2q}{\epsilon_{\infty}}\big(R^s_{1}(\theta)+R^s_{2}(\theta)\big)}.
 \label{eq:Wop}
 \end{equation}
 Although the optical mode, $\omega^s_{op}(q,\theta)$, corresponds to a collective excitation in which the electron densities in the two layers fluctuate in-phase and it is independent of the layer separation $d$ for small wave vectors, the acoustic plasmon mode, $\omega^s_{ac}(q,\theta)$, accounts for an out-of-phase oscillation of the carriers in the two layers and depends on $d$. These branches are influenced by the strain through the strain-dependent orientation factor of layers, $R^s_1(\theta)$ and $R^s_2(\theta)$. While these factors are combined together additively for the optical plasmon modes, they appear as a reduced form, similar to a reduced mass in the acoustic plasmon\cite{Rodin:prb15}. In Fig. \ref{fig:3}, these modes have been illustrated for a few uniaxial strain values along the armchair and zigzag directions in phosphorene double-layer sandwiched by \BN \ at an electron density $n= 1\times10^{13}$ cm$^{-2}$ and with a separation distance $d =5$ nm. When strain is applied to a double-layer system, both the optical and acoustic plasmon modes decrease along the armchair direction. However, these plasmon branches exhibit a long-wavelength behavior similar to the zigzag direction of plasmon dispersion in monolayer phosphorene under uniaxial strain.
 \\
      \begin{figure*}[ht]
      	\includegraphics[width=18.0cm]{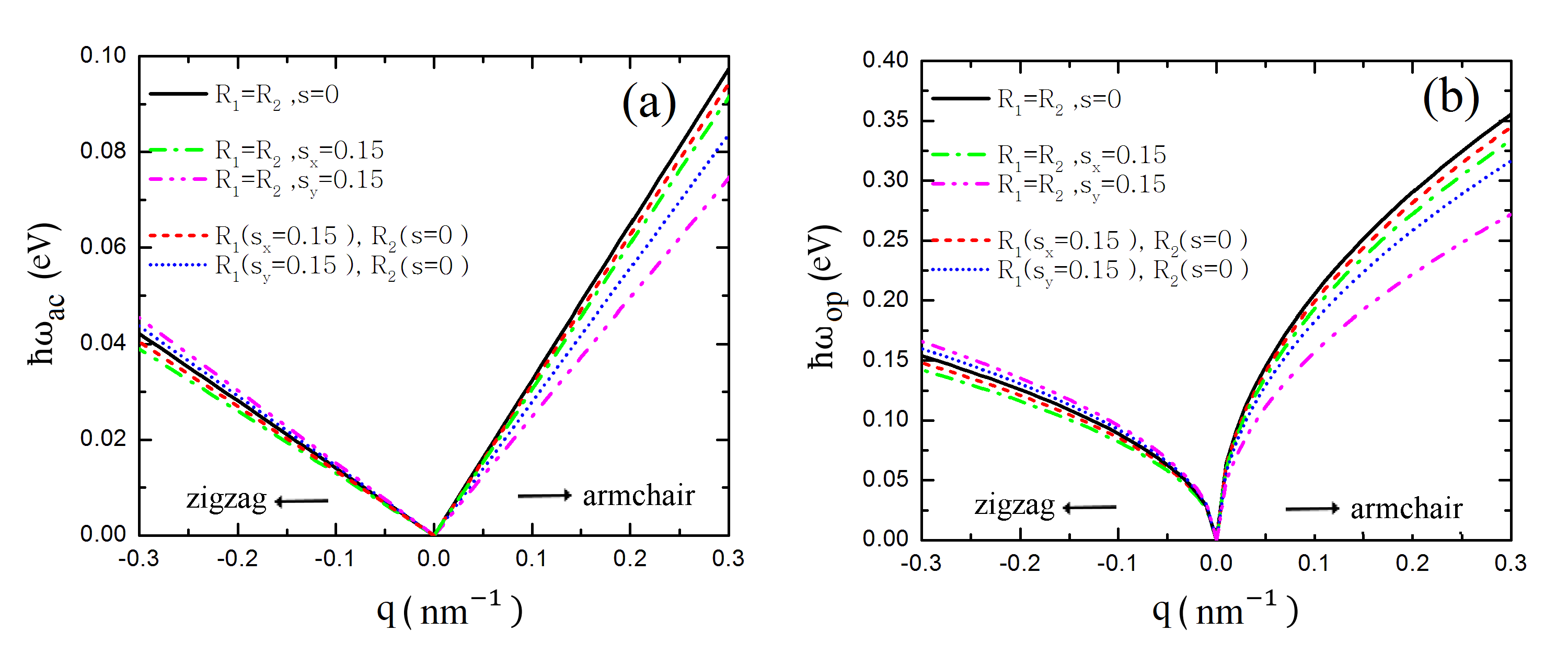}
      	\caption{(a) Acoustic and (b) optical plasmon dispersions along armchair and zigzag directions in a bilayer phosphorene. Uniaxial tensile strain is applied along the $x-$ and $y-$axis with $n= 1\times10^{13}$cm$^{-2}$ and $d =0.5$ nm. The structure is described by the effective dielectric constant of $\sim$ 2 (\BN\  as substrate and air as spacer).}
      	\label{fig:bilayer}
      \end{figure*}     
 It is appealing to model a real bilayer phosphorene by reducing the separation in a double-layer structure down to $d = 0.5$ nm. Therefore, we use the general low-energy model of anisotropic double-layer systems (see Eq. \ref{eq:energy}) in which the inter-layer hopping of electrons is not considered. The plasmon modes of the relaxed and strained bilayer phosphorene are shown in Fig. \ref{fig:bilayer} where an effective dielectric constant of $\sim$ 2 (for a common \BN\ substrate and air spacer) and an inter-layer distance of $d = 0.5$ nm are considered.  More interestingly, due to the weak vdW interaction between phosphorene layers, the plasmon modes of our suggested bilayer phosphorene with no strain are consistent with the results have been previously obtained \cite{Jin:prb15} in the long-wavelength limit. Plasmon modes in Ref.{$\cite{Jin:prb15}$}, were studied by using a TB model for a real bilayer phosphorene in which the inter-layer hopping of electrons was allowed. As expected, the acoustic modes in the bilayer structure are much weaker as compared to the acoustic modes of the double-layer system because of the shorter distance between the two layers. Also, under uniaxial strain, the behavior of the optical and acoustic modes of bilayer system is similar to the case of double-layer phosphorene along both the armchair and zigzag directions. For the strained bilayer structure, we investigate two different cases: \textit{i}) the strain is equally applied to the two layers, $R_{1}(s)=R_{2}(s)$, and \textit{ii}) the strain is applied only to one of the layers, $R_{1}(s\neq 0)$ and $R_{2}(s=0)$. The qualitative difference between these two cases is apparent in Fig. \ref{fig:bilayer}. As it can be observed, the effect of strain on the long-wavelength plasmon energies in case \textit{i} is stronger than the case \textit{ii}.  
              
\section{Transresistivity in a double-layer phosphorene under strain} 
\label{theory3}   
In this section, we study the transresistivity in a double-layer phosphorene under strain. It is well-known that plasmon modes contribute strongly in momentum transfer phenomenon in Coulomb coupled electron gas systems \cite{Flensberg:prb95,Narozhny:prb12}.
The transresistivity matrix for an anisotropic double-layer system which has recently been proposed \cite{Samira:drag2016} is given by\\ 
\\
  \begin{equation}    
     \rho^s_{\alpha\beta}=\dfrac{{\hbar^2}}{(2\pi)^3e^2n_1n_2k_BT}\int\mathrm{d}q\int_0^{\infty}\mathrm{d}\omega F^s_{\alpha\beta}(q,\omega ,T),\\\label{17}
     \end{equation}         
      where $n_1$ and $n_2$ are the electron densities in layers 1 and 2, $\alpha$ and $\beta$ denote the Cartesian coordinates ($x$ and $y$) and $F^s_{\alpha\beta}(q,\omega,T)$ is defined as:\\
     \\
\begin{widetext}
	\begin{equation}
	F^s_{\alpha \beta}({q},\omega,T)=\int_{0}^{2\pi}d\theta \psi_{\alpha \beta}(\theta,\tau_{1},\tau_{2}) \frac{q^{3}}{\sinh^{2}( \hbar \omega/2 k_{B}T)} |U^s_{12}(q,\omega,T;\theta,\tau_{1},\tau_{2})|^{2}\Im\Pi^s_{1}(q,\omega,T;\theta,\tau_{1})\Im{\Pi^s_{2}(q,\omega,T;\theta,\tau_{2})},
	\label{eq18}
	\end{equation}
\end{widetext}                       
     with $\psi^{\alpha\beta}$ given by\\
     \begin{equation}
     \psi^{\alpha\beta}(\theta,\tau_1,\tau_2)=\left\{ \begin{array}{rl}
     \cos(\theta-\tau_1)\cos(\theta-\tau_2),&\alpha=\beta=x\\
     \sin(\theta-\tau_1)\sin(\theta-\tau_2),&\alpha=\beta=y\\
     \cos(\theta-\tau_1)\sin(\theta-\tau_2),&\alpha=x,\beta=y .\\
    \end{array} \right.\label{19}
    \end{equation}
    
     \begin{figure*}[ht]
     	\includegraphics[width=18.0cm]{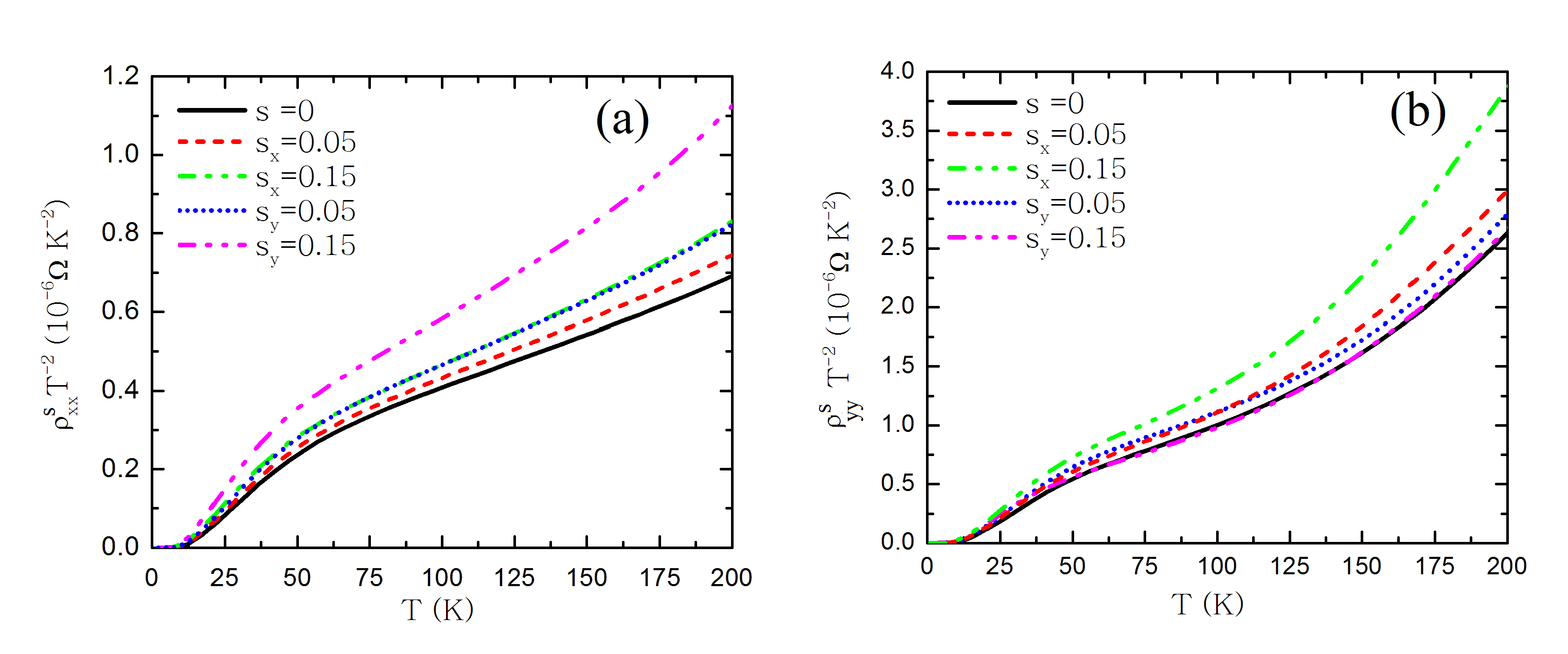}
     	\caption{Scaled drag resistivity for a few values of tensile uniaxial strain along (a) armchair and (b) zigzag directions in a double-layer phosphorene sandwiched by \BN \ with $d = 5$ nm.}
     	\label{fig:5}
     \end{figure*} 
    
    The strain-dependent dynamically screened inter-layer potential $U^s_{12}(q,\omega,T;\theta,\tau_{1},\tau_{2})=U^{s}_{12}(\mathbf{q},\omega)$ can be obtained from\cite{Badalyan:prb12}:
    \begin{equation}
    U^s_{12}(\mathbf{q},\omega)=\frac{V_{12}(q)}{det|\epsilon^s_{12}(\mathbf{q},\omega)|},
    \label{20}
    \end{equation}               
 where $det|\epsilon^s_{ij}(\mathbf{q},\omega)|$ is the determinant of the general dielectric matrix, Eq. (\ref{eq:epsil}). In Fig. \ref{fig:5}, we show the diagonal elements of the transresistivity tensor calculated within the RPA versus temperature for two parallel aligned phosphorene monolayers sandwiched by \BN\ layers and separated by a distance of $d = 5$ nm. It can be seen that the values of $\rho^s_{yy}$ is larger than $\rho^s_{xx}$ in the studied range of temperature. This is a consequence of a higher electron effective mass (Fig. \ref{fig:1}) which results in lower excitation energies for the acoustic and optical branches (Fig. \ref{fig:3}), and thus more contribution of plasmon modes to the momentum transfer phenomenon. Also, as a general result, the transresistivity is mainly influenced by the strain in such a way that it increases by increasing the applied strain. According to the calculations shown in Fig. \ref{fig:5}, the effect of uniaxial strain on the transresistivity is stronger at higher temperatures. In the case of $\rho^s_{xx}$, the drag resistivity shows a neat correlation with the plasmon modes when $s_x$ or $s_y$ is applied. On the other hand, although, the unstrained $\rho^s_{yy}$ has still smaller values than the strained one, it does not really follow the trend of the plasmon modes with applied $s_y$. We think in this case, the effect of applied strain on the single-particle contribution to the drag resistivity dominates the strain effect on the plasmons, resulting in different behavior.
 \\
 \section{conclusion} \label{conclusion}
In this paper, we have studied the effect of an applied uniaxial tensile strain on the plasmon dispersion of monolayer, bilayer and double-layer phosphorene structures.
 As a consequence of anisotropic energy band, the changes of plasmon dispersions are different along the armchair and zigzag directions and strongly depend on the direction of the applied uniaxial strain. Also, in two-layer phosphorene systems, it was shown that the strain-dependent orientation factor of layers control the variations of plasmon energies. In addition, we have obtained that while the behavior of the optical and acoustic modes are similar in both bilayer and double-layer phosphorene, plasmons along the armchair direction are more affected by strain. Moreover, for the strained bilayer structure, two different cases have been investigated: \textit{i}) the strain is equally applied to the two layers and \textit{ii}) the strain is applied only to one of the layers. We have found that the effect of strain on the long-wavelength plasmon modes in case \textit{i} is stronger than the case \textit{ii}. Finally, the diagonal elements of the transresistivity tensor have been calculated within the RPA for a double-layer phosphorene in which two parallel aligned phosphorene monolayers are under strain and sandwiched by \BN. The results have been suggested that the changes in the plasmonic excitations, due to the applied strain, are mainly responsible for the predicted behaviors of the drag resistivity.

 \end{document}